# Scenarios for the Deployment of Automated Vehicles in Europe

[a]Louison Duboz, Ioan Cristinel Raileanu[b], Jette Krause[c], Ana Norman-López[d], Matthias Weitzel[d], Biagio Ciuffo[c]

[a] European Commission Joint Research Centre, Ixelles, 1050, Belgium
[b] Independent Researcher, 077042, Chiajna, Romania
[c] European Commission Joint Research Centre, Ispra, 2107, Italy
[d] European Commission Joint Research Centre, 41092 Sevilla, Spain

**Comment**

This article is submitted to the journal Transportation Research Interdisciplinary Perspectives (TRIPS), from the family of journals Transport Research.

**Acknowledgements**

This research has been funded by the Joint Research Centre of the European Commission. The views expressed are purely those of the authors and may not, under any circumstances, be regarded as an official position of the European Commission.

**Abstract**

The deployment of Automated Vehicles (AVs) is expected to address road transport externalities (e.g., safety, traffic, environmental impact, etc.). For this reason, a legal framework for their large-scale market introduction and deployment is currently being developed in the European Union. Despite the first steps towards road transport automation, the timeline for full automation and its potential economic benefits remains uncertain. The aim of this paper is twofold. First, it presents a methodological framework to determine deployment pathways of the five different levels of automation in EU27+UK to 2050 under three scenarios (i.e., slow, medium baseline and fast) focusing on passenger vehicles. Second, it proposes an assessment of the economic impact of AVs through the calculation of the value-added. The method to define assumptions and uptake trajectories involves a comprehensive literature review, expert interviews, and a model to forecast the new registrations of different levels of automation. In this way, the interviews provided insights that complemented the literature and informed the design of assumptions and deployment trajectories. The added-value assessment shows additional economic activity due to the introduction of automated technologies in all uptake scenarios.

# 1. INTRODUCTION

Automated Vehicles (AVs) are expected to transform road transport, offering numerous benefits, including increased safety, reduced traffic, lower environmental impact, and improved accessibility for various user groups, such as older persons, people with disabilities, and young people (Alonso et al., 2019). However, its introduction faces multiple challenges related to technological and non-technological aspects (Hussain & Zeadally, 2019), for instance, the speed of technological development, test results and regulatory approval, consumers' travel behaviour, quality and affordability of vehicles and services (e.g., for shared cases) and public policies (Litman, 2023).

Although the concept of AV has been formalised a decade ago (SAE, 2021), the introduction of advanced automated systems is relatively new. While driving assistance and partial driving automation have already been on the market for years, conditional driving automation, which does not require human intervention under specific conditions starts to be slowly introduced(Autocrypt, 2023). On the other hand, since 2022, in the U.S. (CPUC, 2022), permits allowing robotaxis to transport passengers have been issued in various cities, while this year, China allowed their use in specific cases (e.g., travel to the airport)(Feifei, 2024). To develop AVs, research and industry have dedicated significant resources during the last decade (Mallozzi et al., 2019) and have carried out pilot studies to test technologies and to assess their public acceptance (Othman, 2022). McKinsey (Deichmann et al., 2023) emphasises investment in research and development as one of the primary challenges related to the deployment of levels 4 and 5 of automation. Main challenges include the development of software for enhancing perception, prediction, and decision-making (e.g., dealing with so-called 'edge cases' that are rare complex driving situations), as well as hardware, where the focus lies on the development of sensing technologies, actuation, and performance computing components.

From a regulatory perspective the situation is also complex. If on one side, automated vehicles are part of the European strategy to make transport safer and more efficient, on the other side, completing the regulatory framework to allow them on the roads is challenging as it requires action by many stakeholders.

Considering European strategy documents, in 2018, the European Commission outlined its goals and initiatives to deploy Cooperative, Connected, Automated and Autonomous Mobility (CCAM) in its Communication '*On the road to automated mobility: An EU strategy for mobility of the future*' (European Commission, 2018). In the same year, the European Parliament welcomed the Commission initiative through the resolution "European Strategy on Cooperative Intelligent Transport Systems" (European Parliament, 2018a); followed by the "Autonomous driving in European transport" resolution (European Parliament, 2018b), urging the Commission to prepare the appropriate regulatory framework, ensuring safe operation and providing for a transparent regime governing liability. In 2019, "The European Green Deal" (European Commission, 2019) called for action to roll out cleaner, cheaper and healthier forms of private and public transport, including connected and automated multimodal mobility systems. In 2020, the "Sustainable and Smart Mobility Strategy" (European Commission, 2020) proposed to deploy automated mobility at a large scale, among its solutions and approaches, to reach 90% reduction in emissions within the transport sector by 2050.

In terms of regulations, building on the legal framework developed in Germany in 2020, the European Commission adopted in 2022, the first worldwide legislation concerning the type-approval of the Automated Driving Systems of fully Automated Vehicles (Ciuffo et al., 2024; European Commission, 2022). This Regulation was a fundamental step for the European Union in a context where new technologies are introduced on the market only if there is a regulation defining requirements for their functioning and their assessment. However, for complex technologies like AVs, placing a system on the market does not automatically ensure that it can be deployed, as this must be defined in National Traffic Regulations, where the European Commission has no mandate to act on behalf of the European Member States (MSs). As a result, while testing automated driving systems is possible in most Member States, their deployment in providing mobility services is possible only in Germany and France. Recognising that this fragmentation may hinder progresses in this field, during the 2024 High Level Dialogue on Cooperative, Connected and Automated Mobility (CCAM, 2024), EU Member States have signed a pledge to jointly commit to develop what is needed to allow a connected and automated transport future in Europe, leaving hope that when the technology will be ready, a proper regulatory framework can be in place.

Considering the above, though the shift towards road transport automation has begun, the timeline towards full automation remains uncertain. In a literature review article focusing on timelines proposed, Agrawal et al. (Agrawal et al., 2023) highlight significant divergences in academic and non-academic literature. The private sector (e.g., companies and investors), exhibited greater optimism, while the public sector (e.g., policymakers, independent agencies, and researchers), which was noted for its more pessimistic outlook. The study also revealed that, as of September 2022, none of the previous predictions regarding levels 3 to 5 of automation had been met. In addition,

Given the existing uncertainty, making estimates about the future deployment of AVs and their economic impact proves to be a complex task. This challenge was described by Raposo et al. (Alonso Raposo et al., 2021), pointing out the divergence in findings from one study to another and revealing the uncertainty surrounding the potential economic benefits of AV deployment. The authors emphasise the divergences in methodologies and assumptions used in these studies, including differences in the timeframe considered. The review also highlights that assumptions related to the deployment of AVs tend to be excessively optimistic, leading to calculations that overstate the potential economic outcomes.

Considering these challenges, this paper aims at presenting a methodological framework to assess a realistic deployment of passengers AVs in EU27+UK by 2050 under three scenarios (slow, medium, fast) together with the economic impact using value-added assessment. Thus, the ambition of this paper is twofold. First, to design three deployment scenarios and corresponding trajectories through 2050, along with assumptions about market entry, automated package costs, and the share in the package value that is allocated to hardware and software across different levels of automation. The design of three scenarios is motivated by the need to address the uncertainty surrounding the deployment of AVs. The second objective is to assess the economic impact of AVs in Europe by calculating the Value Added (VA) generated by the deployment of these technologies.

The paper is structured as follows: The next section presents the different steps of the methodology employed, followed by the results. Finally, discussion and conclusions are presented.

## 2. METHOD

To design three deployment scenarios and estimate the economic impact of passenger AV deployment in Europe, the study followed the steps presented in Figure 1. First, a systematic literature review of academic and grey sources on the expected market entry, deployment trajectories, estimated costs and requirements for hardware and software of the different levels of automation was conducted. Then, elements from the literature review are used to develop a baseline deployment trajectory using the Bass model of technology diffusion (Bass, 1969). Subsequently, expert interviews were carried out to review and refine the preliminary assumptions for the different scenarios and the deployment baseline trajectory. Based on the baseline trajectory, two antagonist trajectories corresponding to a slow and a fast deployment are developed. Finally, the different trajectories are used to assess the economic impact of AVs in the EU27+UK by calculating the Value Added (VA) generated by the deployment of the technology.

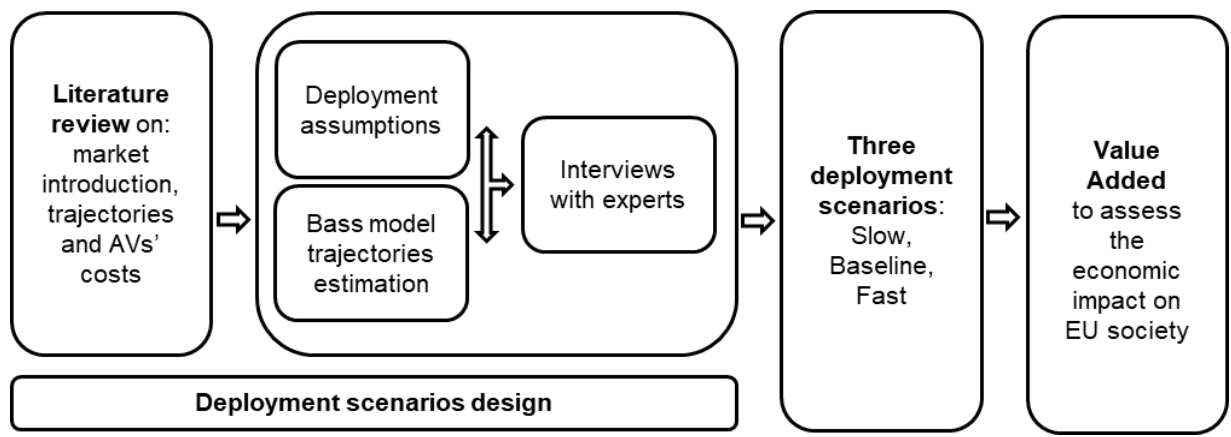

**Figure 1. Structure of the study**

Source: own elaboration.

## 2.1. Literature review

The first step of the analysis consisted of a systematic literature review (Carrera-Rivera et al., 2022) (LR) of academic papers published in peer-reviewed journals and grey literature sources (e.g., government and consulting companies' reports, market reports, articles, press releases, and working papers). The systematic LR focussed on collecting sources that provided details on the expected year of the market availability for various levels of automation, information regarding the specific deployment trajectories, costs of automation technology and an indication of the need for hardware and software. This search was mainly related to levels 3 to 5 of automation technology, as levels 1 and 2 have already been on the market for some years now.

For the scope of this research, the definition of the different levels of automation is based on the SAE International (SAE, 2021) classification, which recognizes six different levels of automated on-road motor vehicles. From levels 0 to 2, the features included in the vehicle support the drivers, while from levels 3 to 5, they can take over the dynamic driving task under specific conditions (level 3) or under all conditions (levels 4 and 5). For the highest three levels of automation, the driver becomes a passenger of the vehicle when the features are engaged.

Academic sources were explored using the Scopus database and Google Scholar, while the Google search engine was used for the grey literature. The search keywords were the terms "Automated /autonomous Vehicles" coupled with the terms "market availability", "year of market introduction", "deployment trajectories", "technology costs /price", "hardware and software needs and costs /price". Before the in-depth analysis of the different sources, a first scope check was performed on the abstracts. The second scope check aimed at the full text.

## 2.2. Development of deployment trajectories

The Bass model of technology diffusion (Bass, 1969) was used to develop the trajectories depicting the expected deployment of the different levels of automation. This model aims at estimating the diffusion of innovative products or technologies. The choice was motivated by the uncertainty surrounding the diffusion of automated technologies, e.g., the contradictory information found in the literature in terms of historical data about low automation levels (levels 1 and 2) and the diverging views for the subsequent levels of automation.

The Bass model has been widely used to forecast the diffusion of innovative products or technologies in various industries and markets (Meade & Islam, 2006). Related to the transport sector, the model was used in multiple studies focusing on electric vehicles (Becker et al., 2020; Bitencourt et al., 2021; Ismail & Abu, 2013; J. H. Lee et al., 2019; Zheng et al., 2020). Massiani and Gohs (Massiani & Gohs, 2015) reviewed the model used in the automotive industry and estimated the Bass parameters' values based on new vehicle registration data in Germany. The model was also used in stock management (Grasman & Kornelis, 2019), and to predict fuel cell vehicle ownership in China (Xian et al., 2022).

To forecast the diffusion of innovative technology, the Bass model builds on three specific parameters: 1) *p,* known as the coefficient of innovation, denotes the consumers buying a novel technology on their own initiative without being influenced by other buyers; 2) *q* is the coefficient of imitation and represents consumers buying the technology due to the influence of previous buyers; and 3) $\overline{N}$ is the market potential of the specific technology and identifies the total number of buyers that will eventually purchase the product over the years.

For estimating the deployment trajectories, the following model equation was used:

$$n(t) = p\overline{N} + (q-p)N(t) - \frac{q}{\overline{N}}[N(t)]^2$$

where $n(t)$ is the number of customers at time *t* and *N(t)* the cumulative number of customers that have already adopted the specific level of automation by time *t*.

The method was used to develop the uptake trajectories for automation level (i.e., 2 to 5) and market uptake scenario (baseline, slow and fast deployment scenarios) in this paper. The three parameters in the model were set and calibrated exogenously (Appendices A and B) present the values used for estimations). This approach was chosen due to the lack of data points allowing an estimation based on historic adoption data of certain levels of automation (Lilien & Rangaswamy, 2000). Specifically, the coefficient of innovation *p* was set to *0.002* for all levels of automation and in all scenarios, based on the values found in the literature that reflected the adoption of new automotive technology in the European contexts (Jensen et al., 2017; Massiani & Gohs, 2015). The q values were calibrated to reach specific values of deployment for the automated technology levels at a given point in time, taken from the literature (see, e.g., Table 5 for the fixed points chosen for the baseline scenario). The market potential $\overline{N}$ for each level of automation was estimated considering the total number of new vehicle registrations in the specific period in which a given level of automation was assumed to be on the market. New vehicle registrations in the EU27+UK until 2050 were taken from the EU Reference scenario 2020[1] (De Vita et al., 2021). The trajectory calibrations were performed in Microsoft Excel 2019.

For this study, the development of the trajectories followed these steps. First, a baseline trajectory was created due to the lack of comprehensive information in the literature necessary to build all three trajectories directly. This baseline trajectory was then presented to various experts (see the next section for more details) who provided their opinions. Based on this input, the slow and fast trajectories were developed, designed to offer contrasting alternatives to the baseline one.

## 2.3. Expert Interviews

Expert interviews were conducted to validate and improve the assumptions derived from the literature review and the preliminary baseline trajectory. The choice of conducting interviews was motivated by the limited amount of information available on the subject and the eclectic nature of the information gathered. The expert interview method was chosen to benefit from experts' technical, process and interpretative knowledge in their area of expertise (Bogner et al., 2009).

The experts were selected based on their knowledge related to automated technologies from a technical or market perspective. To identify them, contacts were gathered from the relevant literature on the subject and through the research team's network built across different projects in the field. In addition, one of the experts was suggested as a relevant source of information while conducting another interview. After this selection, email invitations were sent to twelve experts, out of which seven agreed to be interviewed.

**Table 1. Detailed information on the interviews**

| ID | Organisation type | Organisation's field of operation | Experience related to AVs |
|---|---|---|---|
| **Expert 1** | Start-up | Development of electronics and robotic solutions. | 5 years of experience as chief of operations. |
| **Expert 2** | Start-up | Development of software for autonomous vehicles. | Longstanding experience in designing and working with advanced driving assistance technology, and system development for camera and Lidar technologies. |

---

[1] https://energy.ec.europa.eu/data-and-analysis/energy-modelling/eu-reference-scenario-2020_en

| | | | |
|---|---|---|---|
| **Expert 3** | Trade association | Represents the automotive sector, and produces studies related to this sector. | 8 years of experience in areas related to connected and automated vehicles, electric vehicles, hydrogen fuel cell vehicles, mobility as a service, etc. |
| **Expert 4** | Research organisation | Provides independent studies on technological solutions. | 30 years of experience in areas related to automated driving technology. |
| **Expert 5** | Transport company | Provides end-to-end solutions for electric and autonomous shipping. | 4 years of experience in the system engineering and software area with a particular focus on aspects related to safety. |
| **Expert 6** | Start-up | Development of electronics and robotic solutions. | 6 years of experience in autonomous navigation and its implication for the specific vehicle operated. |
| **Expert 7** | Original equipment manufacturer | Manufacture of vehicles. | Longstanding experience in areas related to driver assistance technology, functional safety, or safe functions for electro-mobility. |

The interviews were conducted online using meeting platforms (i.e., Skype, Teams, Zoom, Google Meet) and lasted between 25 and 40 minutes each. A short interview setting was preferred as it was easier to schedule it with the different experts.

Each interview was structured in five sections. First, the experts were asked to present their understanding and previous experience with AV technologies. The second section focused on the year of market introduction of the different levels of automation from levels 3 to 5. The third section investigated the experts' agreement with the initial baseline deployment trajectory. The fourth section discussed the production costs of the different levels of automation at mass market introduction (defined at 10% of market share). The last part explored the views of the experts regarding the share of hardware and software in the cost of autonomous technology from level 3 up to level 5.

Seven experts from five different European Union Member States and one from a non-EU country were interviewed during May and June 2023. Table 2 provides professional background information on the experts. The information displayed is limited with respect to the privacy statement shared with the experts. Four of them were working for companies that develop technologies related to vehicle automation, two for organisations conducting studies in the field of transport and one for an original equipment manufacturer. Five of the experts had more than 5 years of experience in the field of vehicle automation, while the minimum experience was 4 years.

The interviews were not recorded, but the interviewer took notes during the discussions. After the interview, a summary of the discussions was prepared and sent to the experts to confirm their views and add, if necessary, any information not previously expressed. The summaries were then analysed by applying a content analysis method (Neuendorf, 2018), to draw insights that can support the improvement of the assumptions and trajectories. Specifically, categories were created following the questions of the script and according to the different levels of automation.

## 2.4. Calculation of Value Added

To assess the economic impact on the EU27+UK economy of the deployment of automated passenger cars through 2050, the value added (VA) arising from the production of additional components was calculated. The calculation of the VA follows the approach developed in Alonso et al. (Alonso Raposo et al., 2021) and is based on the different trajectories and assumptions previously developed in this study. Total value added per deployment scenario results as the sum over all automation levels of the costs of the technology times the number of new vehicles registered with the respective automation level.

The development of production costs for each level of automation over time is determined assuming a learning rate approach with a 20% cost reduction for each doubling of production volume. Starting from the assumed production costs in the year of mass market entry of a technology level, defined to be the point in time when the respective automation level reaches a 10% market share, the learning approach is applied both forwards (from the mass market entry year onwards, stopping at a minimum cost of 30% of mass market entry costs) and backwards (higher technology costs before reaching mass market levels). A 50% markup is applied to cover additional costs on top of production costs. The analysis assumes domestic production within the EU27+UK, or balanced import-export for all components of connected and autonomous vehicles.

As regards production volumes, it is assumed that new registrations of passenger cars in the EU27+UK follow the trend of the EU reference scenario 2020 (De Vita et al., 2021).

## 3. RESULTS

In this section, results are presented, starting from the literature review and preliminary uptake trajectories per automation level and scenario, then describing the outcomes of the expert interviews, followed by the revised final trajectories for all scenarios and the results of the value-added calculations.

## 3.1. Literature Review and preliminary deployment scenarios

Main parameters based on the literature review are specified below. They cover the year in which each automation level becomes available on the market, fixed points for the uptake trajectories, costs of automation and their distribution over software and hardware.

A total of twenty-eight sources were identified as providing valuable details for the current paper. Table 1 lists these sources and the areas of our study that they can inform. It can be observed that 11 originate from academic journals, 9 are reports developed by public and private institutions, and 8 are media articles. Notably, the focus of the different sources is on market availability and costs, while trajectories are less investigated. Furthermore, only one source addresses the share in the automation package value that is allocated to hardware and software.

**Table 2. Sources identified in the LR supporting the current analysis**

| References | SAE Levels | Sources typology[1] | Topics covered | | | |
|---|---|---|---|---|---|---|
| | | | Market availability | Trajectories | Costs | Hardware software allocation |
| (Agrawal et al., 2023) | L3, 4 and 5 | AP | + | | | |
| (Becker et al., 2020) | L4/5 | AP | | | + | |
| (Brooke, 2020) | L2, 3 and 4 | MA | | | + | |
| (Catapult, 2021) | L3,4/5 | R | + | + | + | + |
| (Compostella et al., 2020) | L4/5 | AP | | | + | |
| (Concas et al., 2019) | L4 and 5 | R | + | | | |
| (Cusumano, 2020) | L4 and 5 | MA | + | | | |
| (Douma et al., 2019) | L4 and 5 | R | + | | | |
| (Elvik, 2020) | L3 and 4 | AP | | | + | |
| (Groshen et al., 2019) | L3, 4 and 5 | R | + | + | | |
| (Kyriakidis et al., 2015) | L4 | AP | + | | | |
| (Lari et al., n.d.) | L4 | AP | + | | | |
| (Leonard et al., n.d.) | L3, 4 and 5 | R | + | | | |
| (Litman, 2023) | L4 and 5 | R | + | | | |
| (Kersten et al., 2021) | L4 and 5 | MA | + | | | |
| (Deichmann et al., 2023) | L3 and 4 | MA | + | + | | |
| (Mersky, 2021) | L3 and 4 | R | + | | | |
| (Mishra et al., 2021) | L4 and 5 | R | + | | | |
| (Quarles et al., 2021) | L4/5 | AP | + | | | |
| (Shirokinsky et al., n.d.) | L3,4/5 | MA | + | + | | |
| (Simons et al., 2018) | L4/5 | AP | + | | | |
| (Singh et al., 2022) | L5 | AP | | | + | |
| (SMMT, n.d.) | L3, 4 and 5 | R | + | + | + | |
| (Tirachini & Antoniou, 2020) | L5 | AP | | | + | |
| (Wadud, 2017) | L5 | AP | | | + | |
| (Wadud & Mattioli, 2021) | L5 | AP | | | + | |
| (Reiner et al., 2015) | L4 and 5 | MA | | + | + | |
| (Yano Research, n.d.) | L3 and 4 | MA | + | + | | |

[1] *Acronyms: MA – Media Article, AP – Academic Paper, R – Report*

Source: own elaboration.

### 3.1.1. Market availability

Considering the market availability of levels 4 and 5 of automation, the sources analysed provide a wide range of estimations. Sources reviewed align with Agrawal et al. (Agrawal et al., 2023) and highlight a lack of agreement regarding their estimations. Optimistic sources (Concas et al., 2019; Kyriakidis et al., 2015; Simons et al., 2018) anticipate the availability of Level 4 vehicles by 2025 and those of Level 5 by mid-2030. In a survey of executives from automotive, transportation, and software companies, McKinsey (Kersten et al., 2021) indicated that Level 4 could be available for specific driving tasks (i.e., highway pilot) by 2025, while Level 4/5 vehicles (e.g., robotaxi in urban areas) could be available by 2030 in the European context. Other estimates place the initial commercial availability of AVs at a high price premium in the decade 2030-40 and predict an extensive deployment after 2045 (Litman, 2023). On the other side, sceptical views (Cusumano, 2020; Mersky, 2021; Mishra et al., 2021; Quarles et al., 2021) consider Level 5 vehicles as a distant goal unlikely to happen before 2050. From the sources analysed, it can be observed that studies published before 2019 were more optimistic about the deployment and availability of highly automated vehicles (Levels 4 and 5), while recent studies are more conservative.

To account for the lack of agreement on the timeline for introducing higher levels of automation, the current paper considers three scenarios for the market availability of AVs. Table 3 presents the preliminary estimates regarding the years when the specific levels of automation could reach 1% of the new vehicle registrations in the EU27+UK. The references presenting estimations aligning with the different scenarios are detailed in appendix C.

**Table 3. Preliminary assumptions for market availability**

| Level of automation | Year* | | |
|---|---|---|---|
| | SLOW | BASELINE | FAST |
| **Conditional automation – Level 3** | 2030 | 2025 | Before 2025 |
| **High automation – Level 4** | 2040 | 2035 | 2030 |
| **Full automation – Level 5** | - | 2040 | 2035 |

*The year when the specific level of automation could represent 1% of the new vehicles' registrations in the EU
Source: own elaboration.

### 3.1.2. Trajectories

The annual aggregated data are missing at the EU level regarding new vehicle registrations for the various levels of automation. This section summarises results from the literature review and documents the assumptions made when developing the baseline scenario trajectory.

Level 1 and 2 vehicles have been on the market already since the early 2010s . For the baseline scenario, the estimates from SMMT (SMMT, n.d.) for the United Kingdom market, where Level 1 was found to reach 69% in both 2020 and 2025. Regarding Level 2, the trajectory development was based on Roland Berger's (Shirokinsky et al., n.d.) projection of 39% in 2025. For Level 3, an uptake of 1% of the new passenger vehicle registration in 2025 was considered, and an estimation of 8% in 2030 reported in the central scenario by Catapult (Catapult, 2021) and in Yano Research Institute (Yano Research, n.d.). For Levels 4 and 5, it was assumed that Level 4 will represent 1% of the new passenger vehicles in 2035 and that Level 5 will be 1% of new vehicle registrations in 2040. Table 4 presents the values from the literature used for developing the baseline scenario trajectories.

**Table 4. Uptake of automation levels from literature used as fixed points for fitting the Baseline**

| Level of automation | Source(s) | Values for developing the trajectories |
|---|---|---|
| Level 2 | Roland Berger (Shirokinsky et al., n.d.) | 39% in 2025 |
| Level 3 | Catapult (Catapult, 2021); Yano Research Institute (Yano Research, n.d.) | 8% in 2030 |

Source: own elaboration.

### 3.1.3. Costs of the different levels of automation

Regarding the cost for each level of automation, the literature provides diverse values. Our assumptions on additional production cost to reach the various automation levels are presented in Table 5. For Levels 3, 4 and 5, values were conversed from British pound sterling to Euros at 2022 conversion rate.

**Table 5. Assumptions regarding the additional production costs to reach specific levels of automation at 10% market penetration**

| Level of automation | Source(s) | Additional production cost per passenger vehicles |
|---|---|---|
| **Level 1** | (SMMT, n.d.) | 814 € |
| **Level 2** | (SMMT, n.d.) | 1,628 € |
| **Level 3** | (Catapult, 2021) | 3,579 € |
| **Level 4** | (Catapult, 2021) | 6,301 € |
| **Level 5** | (Wadud, 2017); (Wadud & Mattioli, 2021) | 10,934 € |

Source: own elaboration based on literature.

### 3.1.4. Hardware and software shares of additional costs

Regarding the hardware and software shares in the additional costs to reach given deployment levels, only one source was identified that provides this specific information (Catapult, 2021). The values are presented in Table 6.

**Table 6. Shares of hardware and software in total cost of the automated package**

| Level of automation | Hardware | Software |
|---|---|---|
| Level 3 | 65% | 35% |
| Level 4 and 5 | 50% | 50% |

Source: Catapult (Catapult, 2021).

## 3.2. Preliminary market uptake

Based on the systematic literature review presented in the subsections "Market availability" and "Trajectories", the trajectory based on the baseline scenario assumptions was developed.

To estimate the trajectory, the Bass model was used as explained in the methodology subsection "Development of automation trajectories". The coefficient of innovation (p) was set at 0.002 for all levels of automation, and the values for the coefficient of imitation (q) were calibrated to reach specific deployment levels of the technology, such as a 39% for Level 2 vehicles in 2025 and a value of 8% for the Level 3 vehicles in 2030. Appendix A lists the coefficient values used for developing the trajectories and the estimation timeframes for each level of automation technology.

Figure 2 presents the preliminary trajectory develop considering the baseline scenario assumptions (Table 3). Values represent percentages of new vehicle registrations in the respective year equipped with the specific level of automation technology.

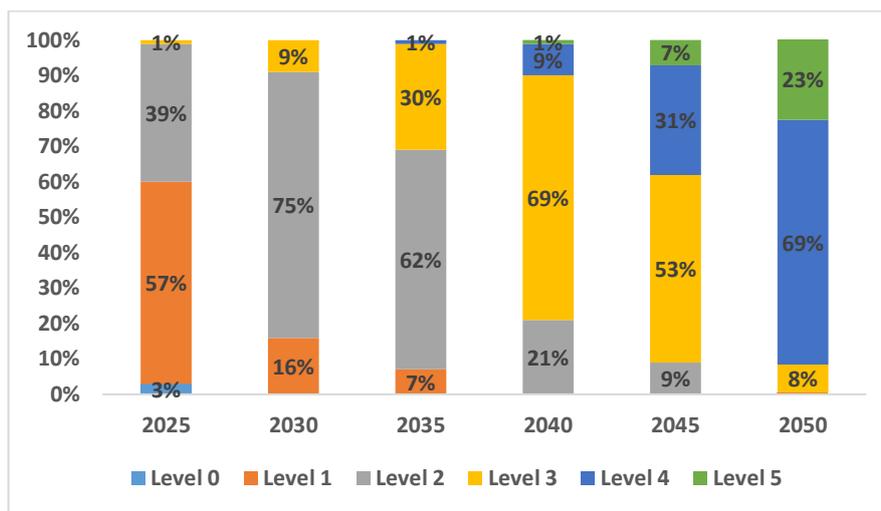

Source: own elaboration.
**Figure 2. BASELINE scenario trajectory for the period 2020 – 2050**

## 3.3. Expert Interviews

The expert interviews reflected on the preliminary assumptions on various aspects of AV deployment, including market introduction, automation trajectories, production costs, and the allocation of resources between software and hardware.

Regarding the year of market introduction for different automation levels, for level 3 of automation, experts highlighted that the technology has been present on the market after 2020. Hence, assuming a 2025 market entry was reasonable given the five-year increments considered in the method. Although, two of them explained that Level 3 could reach the (10%) mass market in the next five years or towards 2030. With respect to Level 4, the expected introduction year among the experts, with estimations varying from 2030 to 2040.

Optimistic experts highlighted that around 2030, this level could be introduced for public transport first (i.e., robot-taxi and shuttles) and as a logistics alternative (i.e., robots, trucks). Level 5 was perceived as "utopic" and still unavailable in 2050 by five experts due to challenges in the operational design domain (ODD), the necessity for all vehicles to be fully automated and the need for appropriate infrastructure. However, one expert shared a more progressive opinion and proposed 2045 as a starting year for Level 5.

Regarding the deployment trajectories, only two experts provided information, briefly, on Level 1 of automation. For Level 2, four experts agreed that this level would stay on the market longer than preliminarily expected; even beyond 2050, according to one of them. They perceived this level as a pragmatic way to deploy AV technology. Indeed, Level 2 can support the development of higher levels of automation by providing valuable transport data gathered by the sensors, in particular traffic flow information and driving data, while simultaneously introducing higher levels of automation to consumers at affordable prices. One expert offered insights on the market share, suggesting that achieving a 40% market share by 2025 was overly ambitious. Instead, they proposed that a more realistic goal would be around 25% or possibly up to 30%. For Level 3, most experts agreed with the preliminary baseline trajectory proposed in Figure 2, although one described it as too optimistic and said that they would not expect 1% of market share for this level before 2030. Most experts agreed that the mass market could be reached by 2028 for Level 3 of automation. Interestingly, two experts highlighted this level as a transition technology towards higher levels of automation, adding that the introduction of Level 4 could reduce the demand for Level 3. With respect to Level 4 and 5 trajectories, opinions were divided. Some experts stated that the trajectories were in line with expectations, while others described them as too optimistic or indicated that the trajectories would depend on the business model adopted and the class of the vehicles.

When considering the hardware and software shares in the additional cost of the technology at mass market entry (defined as 10% of market share), two experts emphasised that the shares might fluctuate over time, evolving with changes in economies of scale. Therefore, predicting this information is challenging. Furthermore, two experts provided data indicating that the balance between software and hardware will shift towards a higher share of software as the level of automation increases. Specifically, while the hardware share is higher for Levels 1 and 2, this ratio changes as the levels of automation increase and more complex software is required.

Regarding the cost estimates of the different levels of automation at mass market introduction, none of the experts provided feedback on Levels 1 and 2. The Level 3 value was perceived as a good cost estimate, although one expert thought that it could be a bit lower. Two experts suggested that the cost of Level 4 should be aligned with that of Level 3, as their costs should be similar, and a third expert agreed with this estimate. When considering Levels 4 and 5, two experts emphasised that the cost should be higher based on their enumeration of the costs of the different components needed for this level of automation. Regarding Level 5 alone, one expert proposed a lower cost (€8000), and another one explained that it was difficult to estimate.

## 3.4. Final deployment scenarios

Based on the expert interviews, the main assumptions underlying the deployment scenarios were revised, and revised trajectories were calibrated.

Descriptions of scenarios envisaged for the deployment of higher levels of automation technology are included in Table 7. These were developed considering a foresight approach as detailed in (Spaniol & Rowland, 2018) and describe briefly the main driving forces that can influence the deployment and adoption in the EU27+UK of highly automated vehicles. It considers the preliminary market availability perspective presented in Table 3 and focuses on aspects such as technology readiness and investment, consumer adoption and the pace of the regulatory framework.

Table 7. Description of scenarios

| SLOW | BASELINE | FAST |
|---|---|---|
| In this scenario, the path of automated technologies is slower due to low investment in research and development, as well as challenges in deployment and adoption of Levels 4 and 5.  The legal framework is fragmented with reduced coordination between the EU level and the Member States. Therefore, there are regulatory inconsistencies and operational challenges for the development and deployment of Levels 4 and 5.  The reduced trust in AV technologies and high safety concerns explain a low demand from consumers for highly automated vehicles. Level 4 vehicles will be available on the market in the year 2040, and Level 5 could be introduced after 2050. | The pace of technology development increases constantly, and the investment in automated technology steadily grows over the next decade, supporting the market introduction and transition to higher levels of automation.  The legal framework keeps pace with the technological advancements and the consumers' acceptance and adoption of AVs. This facilitates the market introduction of highly automated vehicles of Level 4 in 2035 and Level 5 in 2040.  The successful deployment of Level 3 increases consumers' trust and reduces safety concerns regarding AV technology. Therefore, the demand for Levels 4 and 5 is pushed by this previous experience. | The deployment of AV technologies is accelerated thanks to investments in research and development. This trend allows to overcome technological barriers.  The legal framework supports and favours the development and deployment of AVs. Subsidy schemes and various incentives support the purchase and use of Level 4 and 5 vehicles.  Mobility improvements (e.g., reduced number of accidents; enhanced accessibility for various groups of persons) introduced by highly automated vehicles help consumers to overcome their safety and trust concerns. Consumers' acceptance and use strongly supports the demand and deployment of AVs (Levels 4 and 5). |

Source: Own elaboration.

Table 8 presents the revised market introduction years of technology levels, where technology shares reach a minimum penetration of 1%. Compared to the preliminary assumptions, the only change is related to the baseline scenario, as different experts indicated that Level 5 vehicles would not reach the mass market before 2045 or even later.

**Table 8. Final market introduction years, revised based on the input from experts**

| Level of automation | Year* | | |
|---|---|---|---|
| | SLOW | BASELINE | FAST |
| Conditional automation – Level 3 | 2030 | 2025 | Before 2025 |
| High automation – Level 4 | 2040 | 2035 | 2030 |
| Full automation – Level 5 | - | 2045 | 2035 |

*The year when the specific level of automation could represent 1% of the new vehicles' registrations in the EU27+UK

Source: own elaboration.

Figure 3 presents the revised trajectories (compared to the original available in Figure 2). As before, values represent percentages of new vehicle registrations in the respective year for the specific level of automation. Based on the feedback provided by the experts during the interviews, the market potential of Level 2 vehicles was increased to 31.6% compared to the preliminary estimations. The experts' considerations regarding Level 2 translated into estimating a higher market potential for this level compared with the preliminary estimations. This change led to reductions in the market potential of higher automation levels (Levels 3 to 5). It also triggered adjustments in the coefficient of imitation (q) to reach the chosen fixed values from the literature (e.g., 39% for Level 2 in 2025 and 8% for Level 3 in 2030).

In this section, we also present the trajectories developed based on the assumptions considered in the slow and fast scenarios. These trajectories were developed with the Bass model, as described in the methodology section. The value of 0.002 was used for the coefficient of innovation (p), and the values for the coefficient of imitation (q) and the market potential of each level of automation were calibrated based on the values used for developing the baseline trajectories as presented in Annex B.

As observed, Level 2 automation remains on the market longer in the slow scenarios, while it disappears after 2035 in the fast scenario. Notably, Level 2 persists longer in the baseline scenario compared to the preliminary one, aligning with experts' views that emphasize its potential in supporting the development of higher levels of automation by providing valuable transport data. Level 3 disappears by 2045 in the fast scenario due to the rapid adoption of Levels 4 and 5 automations. In contrast, the adoption of higher levels of automation is substantially lower in the baseline scenarios, aligning with most experts' views.

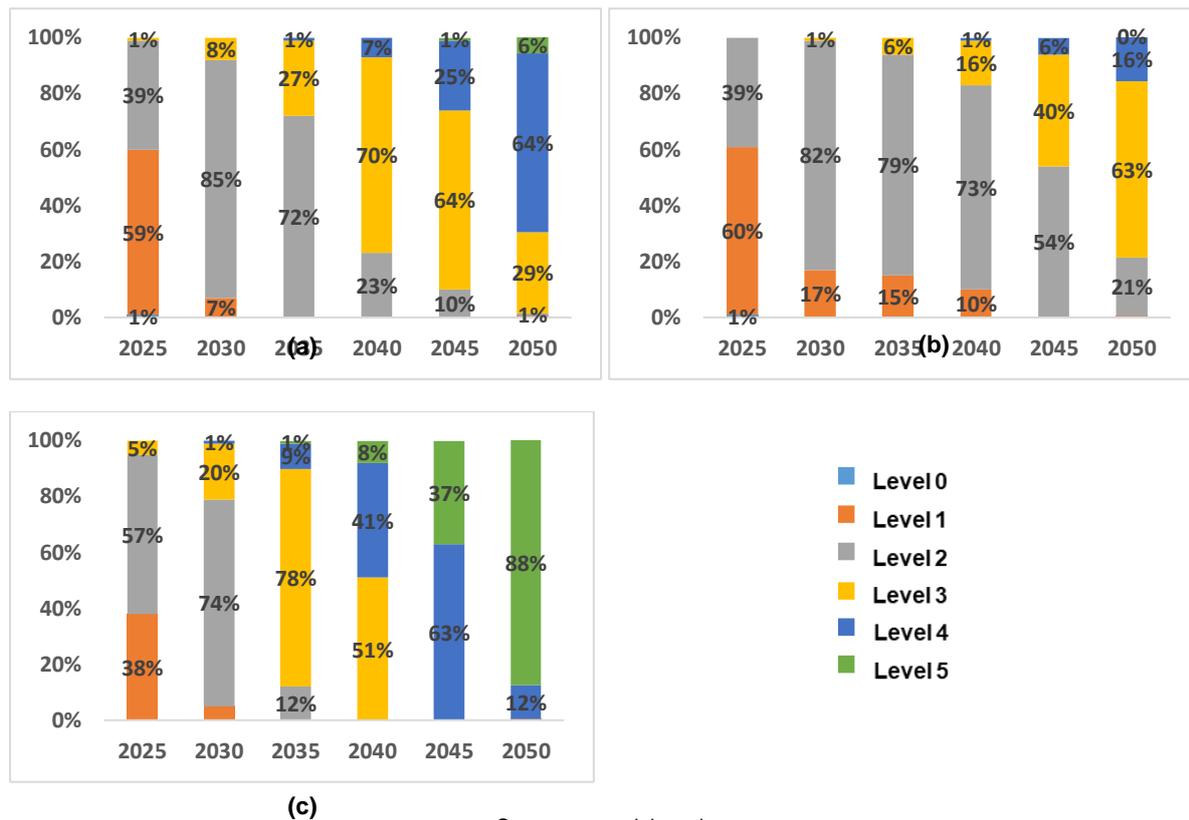

Source: own elaboration.

**Figure 3. Updated market uptake trajectories for the period 2025 – 2050: (a) baseline; (b) slow; (c) fast scenario**

## 3.5. Value Added from passenger car automation

In economic terms, the production of additional software and hardware components for vehicle automation generates additional economic activity and thus increases value added. Based on the production costs per automation level combined with a learning approach and on the assumptions made for new passenger car registrations in the EU27+UK until 2050, value added has been calculated for each of the final automation trajectories developed in this study. In line with the increasing uptake of higher automation levels, value added creation starts at a low level and increases over time. Outcomes for the baseline scenario are shown in Figure 4. As it can be seen, annual value-added increases from around 10 billion Euro in 2020 to 18 billion Euro in 2050. Total value added in the period from 2020 to 2050 amounts to 1.5trillion Euro in the baseline scenario. It ranges from around 950 billion Euro in the slow scenario and 2.6 trillion Euro in the fast uptake scenario. Figures for both the slow and the fast scenario are included in Appendix C.

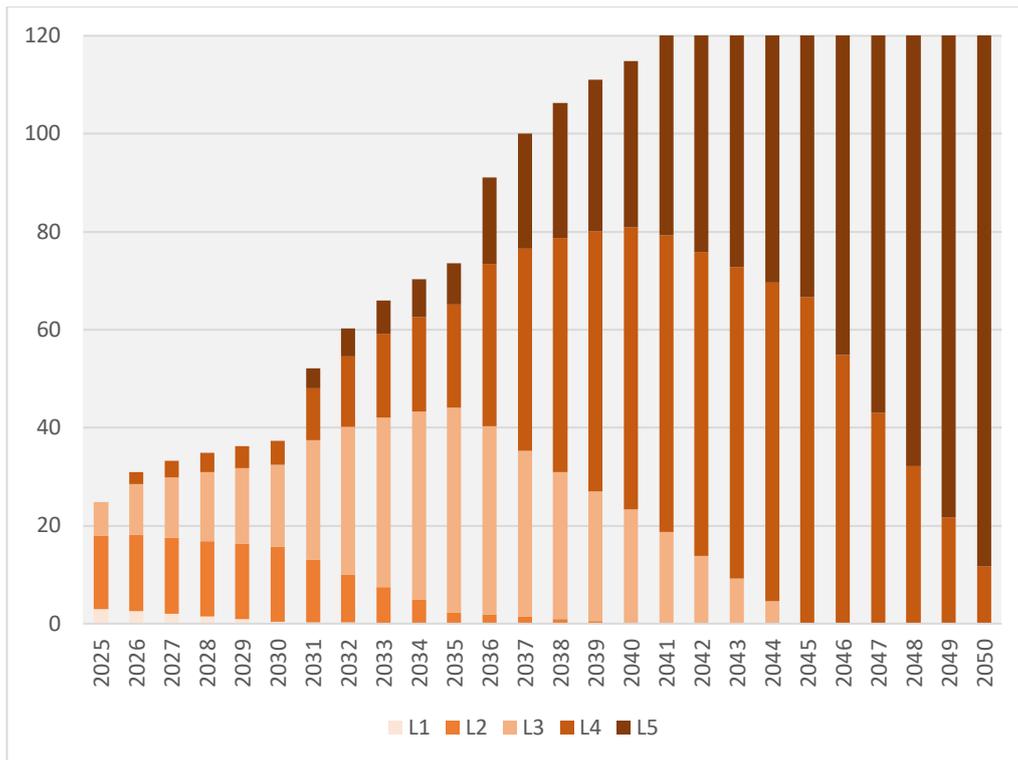

Source: own elaboration.

**Figure 4. Value added results for the baseline automation scenario [billion EUR], per automation level from 2025 to 2050**

Compared to previous results on value added from vehicle automation presented in Alonso et al. (Alonso Raposo et al., 2021), in the present scenarios, maximum possible value added is substantially lower, in particular in the period up to 2040. In the previous calculations, which combined different new vehicle registration trajectories with a set of uptake scenarios for automation, the total additional value added was calculated over the period from 2020 to 2050 and ranged from 825 billion EUR for the low registrations / low automation scenario to 3.6 trillion EUR for the high registrations / high automation combination. The smaller maximum value added presented here is due to the revision of the uptake scenarios for automated vehicles to the more moderate assumptions of maximum uptake presented in this paper, to the lower assumed cost estimates for higher automation levels 3, 4 and 5, and to the fact that all scenarios use the same reference new registration trend, whereas the previous high scenario combination also assumed a stronger growth in new registrations over time. Both the uptake expectations and the cost estimates for high automation levels have been revised downwards significantly compared to previous work, based on literature and expert estimates.

## 4. DISCUSSION AND CONCLUSIONS

While vehicles with low levels of automation have been on the market for about a decade, and Level 3 automation is just emerging in Europe, higher levels of automation (Level 4 and Level 5) are expected to be deployed soon. However, only a few studies cover the deployment of AVs, and these studies often lack information or contradict each other regarding the different aspects this paper has been investigating. This makes it difficult to forecast the impact of automation technologies on the economy.

This paper presents a methodology for future AV deployment projections, which involves qualitative methods such as a comprehensive literature review and interviews with experts, as well as the application of a model to estimate the new registrations of different levels of automation up to 2050. This combined methodological approach has supported the development of deployment scenarios representing the market entry and deployment (i.e., share of new registrations) of automation Level 0 to Level 5 from 2025 to 2050. The final assumptions and trajectories are then used to calculate the value-added due to the additional production of software and hardware.

The value-added results due to deploying additional hardware and software in vehicles show additional economic activity due to the introduction of automated technologies in all the uptake scenarios. The value-added increases with faster uptake and becomes larger as the level of automation increases, demonstrating potential benefits for vehicles or component producers through the introduction of higher levels of automation. However, while value added in vehicle production increases, automation technology presents an extra cost for the buyers and users of vehicles, i.e., private households or companies depending on the transport of people and goods for their production. This could be compensated by productivity increases through less time allocated to driving (Yankelevich et al., 2018). The resulting net economic impacts in different sectors and in the EU-wide total, in terms of GDP and employment, cannot be determined by the present study. Moreover, automation of higher levels can also have significant impacts on transport demand (e.g., by making trips more attractive or goods transport less costly), and could, therefore, trigger impacts on transport and related energy demand as well as congestion. These aspects merit further investigation to provide a full picture.

In the method developed and presented in this article, the lack of information on different aspects of the deployment of AVs in the academic and grey literature was overcome by using both quantitative and qualitative methods. First, the Bass model allowed us to define an uptake trajectory based on data points provided in the literature, while the interviews allowed to improve the baseline trajectory and to refine the scenarios assumptions in general.

More precisely, the Bass model, given its multiple applications in the literature (Ismail & Abu, 2013; Kumar et al., 2022; H. Lee et al., 2014; Massiani & Gohs, 2015) and its identification as a reliable method for demand forecasting (Hong et al., 2016), still has some drawbacks.

These include the assumptions that the factors influencing adoption remain constant over time and that there is no variation among the values of the coefficients used. Other constraints include the omission of several variables that could affect diffusion, such as marketing efforts, changes in market competitive dynamics, subsidies for purchasing vehicles, and regulatory changes over time. Another limiting factor for the current research is the difficulty in accounting for variability among the EU Member States' and UK vehicle markets, as the trajectories in the paper are developed under the assumption that the Bass coefficients reflect the entire EU27+UK vehicle market.

Regarding the interviews, expert knowledge fruitfully complemented scientific and grey literature and modelling approaches by providing new insights on the future deployment of automated vehicles. However, the analysis clearly indicates that uncertainties remain high concerning the deployment of AVs. While the experts had divergent opinions on various aspects, such as market availability, costs, and baseline trajectories, it appeared to be more challenging to provide clear opinions on the shares of hardware and software. To account for this uncertainty, slow and fast scenarios were developed to provide comparison with more extreme cases.

As stated before, this article presents the first step of research aiming at assessing the economic impact of the deployment of AVs on the European Union through 2050. In the next steps, the different trajectories and assumptions presented here will be implemented in the global macroeconomic general equilibrium model, JRC-GEM-E3[2] (Tamba et al., 2022), to assess the overall economic impacts on the European economy in terms of gross domestic product and implications for various sectors in terms of production levels and employment.

---

[2] https://joint-research-centre.ec.europa.eu/gem-e3_en

# APPENDIX A- COEFFICIENT ASSUMPTIONS FOR THE PRELIMINARY BASS MODEL ESTIMATIONS

Table 9 presents the preliminary coefficient assumptions used to calculate the baseline trajectory.

**Table 9. Summary of coefficients assumptions for the preliminary scenarios**

| Level of automation | Scenario | p | q | $\overline{N}$ | Estimated period |
|---|---|---|---|---|---|
| **Level 2** | Baseline | 0.002 | 0.325 | 181,040,000 vehicles | 2015 - 2029 |
| **Level 3** | Baseline | 0.002 | 0.3 | 206,769,000 vehicles | 2025 - 2039 |
| **Level 4** | Baseline | 0.002 | 0.3 | 220,949,000 vehicles | 2035 - 2049 |
| **Level 5** | Baseline | 0.002 | 0.3 | 164,194,000 vehicles | 2040 - 2050 |

# APPENDIX B - COEFFICIENT ASSUMPTIONS FOR THE UPDATED BASS MODEL ESTIMATIONS

Table 10 presents the updated coefficient assumptions used to calculate the baseline trajectory after the expert interviews.

Table 10. Summary of coefficients assumptions for the revised scenario

| Level of automation | Scenario | p | q | $\overline{N}$ | Estimated period |
|---|---|---|---|---|---|
| **Level 2** | Slow | 0.002 | 0.26 | 285,823,000 vehicles | 2015 - 2030 |
| | Baseline | 0.002 | 0.285 | 238,186,000 vehicles | 2015 - 2030 |
| | Fast | 0.002 | 0.04 | 163,321,000 vehicles | 2015-2027 |
| **Level 3** | Slow | 0.002 | 0.26 | 146,134,000 vehicles | 2030 - 2050 |
| | Baseline | 0.002 | 0.335 | 143,879,000 vehicles | 2025 - 2041 |
| | Fast | 0.002 | 0.04 | 170,148,000 vehicles | 2023 - 2035 |
| **Level 4** | Slow | 0.002 | 0.26 | 146,134,000 vehicles | 2040 - -2050 |
| | Baseline | 0.002 | 0.335 | 111,026,000 vehicles | 2035- 2050 |
| | Fast | 0.002 | 0.04 | 146,134,000 vehicles | 2030 - 2043 |
| **Level 5** | Slow | - | - | - | - |
| | Baseline | 0.002 | 0.335 | 111,026,000 vehicles | 2045 - 2050 |
| | Fast | 0.002 | 0.04 | 133,231,000 vehicles | 2035 - 2050 |

# APPENDIX C - LITERATURE SOURCES PRESENTING ESTIMATIONS IN LINE WITH THE SCENARIOS CONSIDERED

Table 11 presents references used to build the assumptions of the different scenarios.

**Table 11. Literature sources for scenarios' estimation**

| Scenario | Paper/ report | Estimations on the introduction of Levels 3 to 5 |
|---|---|---|
| **Slow** | Cusumano, (2020) | Level 3 will develop further towards 2030. Level 4 or 5 remains a distant goal. |
| | Quarles et al., (2021) | Alternative scenario AVs representing 1% in 2040. |
| | Mersky, (2021) | Level 3 available by 2030. No info on Levels 4 and 5. |
| | Mishra et al., (2021) | No near timeline for the launch of vehicles with Level 5 capabilities on public roads. |
| | McKinsey, (2023) | Delayed case scenario with L3 estimated at around 4% in 2030 and L4 (highway pilot) available in 2035. |
| **Baseline** | Lari et al., 2015 | 2020s the availability of the AVs and a wider adoption in 2040-2050 |
| | Douma et al., 2019 | 2040-2050 period for availability and wider adoption of vehicles with full self-driving capabilities |
| | Leonard et. al., 2020 | Level 4 towards 2030. Level 5 beyond 2040. |
| | Catapult, (2021) | Central scenario that estimates Level 3 to represent approximately 2% of the new car sales in 2025. |
| | Litman, 2023 | Level 4 commercially available by 2030 and could reach half of new vehicles by 2045. |
| **Fast** | Kyriakidis et al., (2015) | Survey of general population, median responses regarding the year of deployment were for Level 3 in 2018, Level 4 in 2025 and Level 5 in 2030. |
| | Simons et. al., 2018 | Level 4 or 5 market availability around 2025 could reach at least ¼ of the market by 2035 – 40 and by 2050 all new vehicles will be automated. |
| | Concas et al. 2019 | Level 4 may be available in the mid-2020s to early-2030s. Level 5 expected to have an impact in the mid- to late-2030s. |
| | Yano Research Institute, (2022) | Level 3 estimated at 8% in 2030 and Level 4 to represent around 1% in 2030. |
| | McKinsey, (2023) | Accelerated scenario with L4 around 10% of the vehicles sold in 2030, reaching around 40% by 2035. |

*Source: own elaboration.*

# APPENDIX D - VALUE ADDED RESULTS FOR THE SLOW AND FAST AUTOMATION TECHNOLOGY UPTAKE SCENARIOS

Figures 5 and 6 present the value added resulting from the slow and fast automation technology uptake.

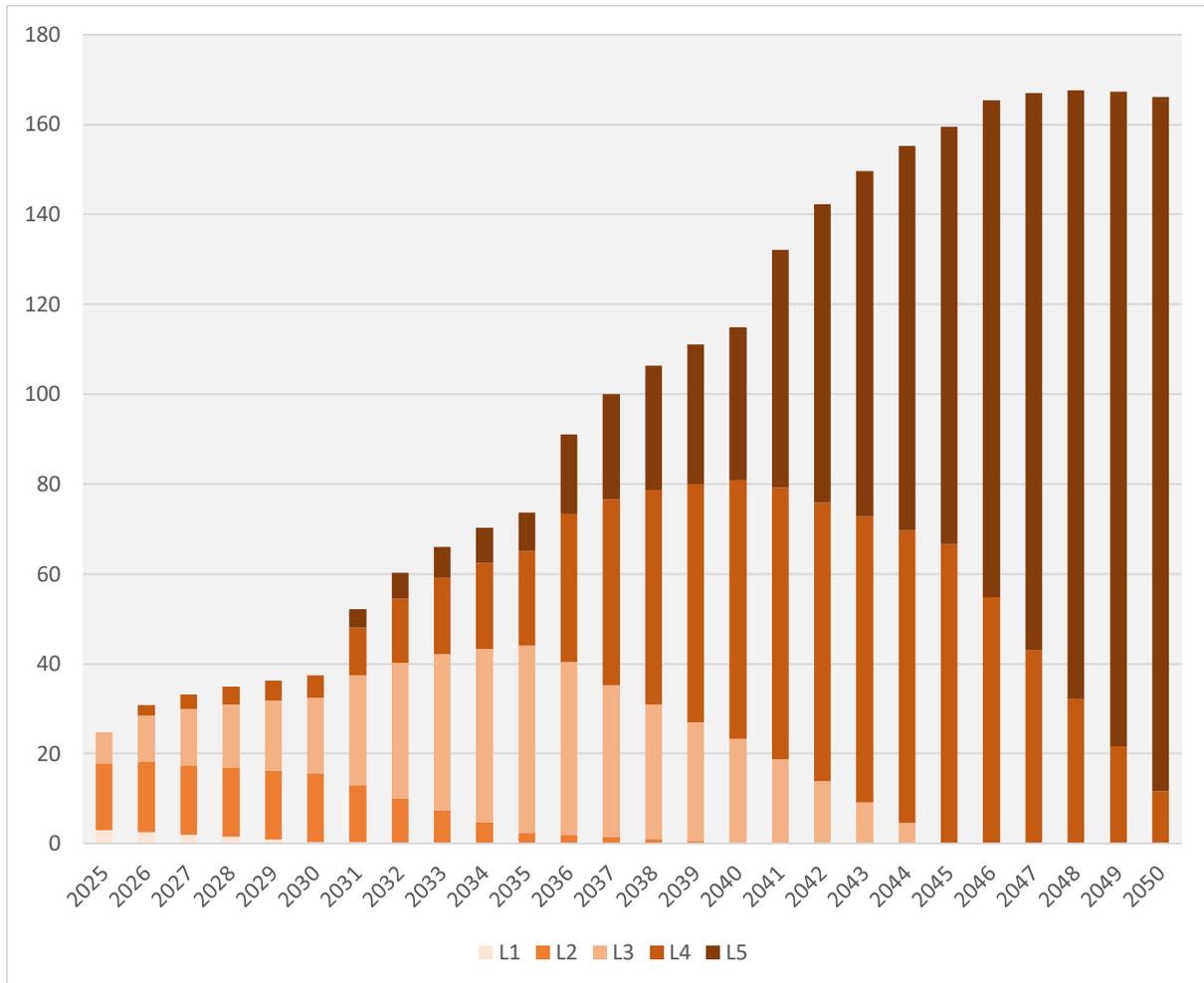

*Source: own elaboration.*

**Figure 5: Value added results for the slow automation scenario [billion EUR], per automation level from 2025 to 2050**

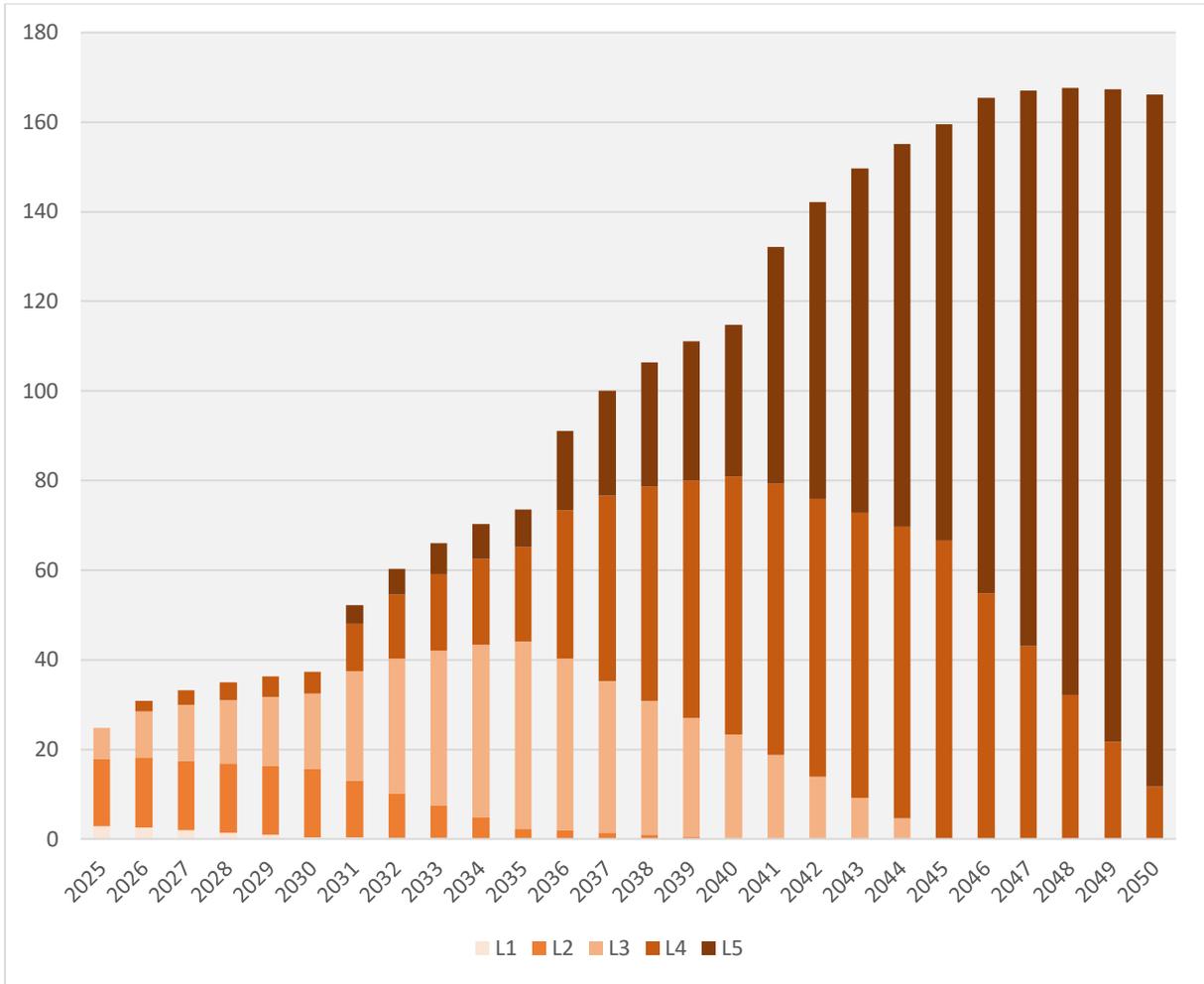

*Source: own elaboration.*

**Figure 6: Value added results for the fast automation scenario [billion EUR], per automation level from 2025 to 2050**